\documentclass{siuro250211}

\usepackage{graphicx} 
\usepackage{float}
\usepackage{amsmath}
\usepackage{amsfonts}
\usepackage{array}
\usepackage{tcolorbox}
\usepackage{tikz}
\usepackage{cleveref}
\usepackage{pythonhighlight}
\usepackage{verbatim}
\usepackage[utf8]{inputenc}
\usepackage{multirow}
\usepackage[backend = biber, style = ieee]{biblatex}
\bibliography{references}

\newcommand{\pt}{\partial_t}
\newcommand{\px}{\partial_x}
\newcommand{\R}{\mathbb{R}}
\DeclareMathOperator*{\argmin}{arg\,min}

\title{Micro-Macro Simulation of Shallow Water Moment Equations}
\author{Vilém Rožek\footnote{vilemrozek@gmail.com}}
\dedication{Project advisors: Julian Koellermeier\footnote{j.koellermeier@rug.nl}, Rik Verbiest\footnote{r.verbiest@rug.nl}}
\date{July 2025}

\begin{document}

\maketitle

\begin{abstract}
    Shallow flows are governed by the Navier-Stokes equations. They are commonly modelled using the shallow water equations, a great simplification of the Navier-Stokes equations, which often yields inaccurate results. For that reason, a model called shallow water moment equations has been developed. It uses more equations and variables than the shallow water equations. While this model is significantly more accurate, it is also computationally more expensive. To speed up computations, the micro-macro method may be used. The micro-macro method switches between two models of varying levels of detail allowing for larger stable time steps.\\ 
    In this paper we formulate the micro-macro method for shallow water moment equations. We perform a theoretical runtime analysis of the method and present a series of results for a dam break test and a wave transport test. The micro-macro method achieves a significant speed-up while retaining a sufficient level of accuracy.
\end{abstract}

\section{Introduction}
\label{sec:1}

In free-surface flows, the term shallow flow refers to situations where the horizontal length scale is much larger than the vertical length scale. Computer models for shallow flows are widely used in weather forecasting, free-surface hydraulics in rivers and channels, assessment of avalanches and landslides, computation of granular transport processes, and modelling coating processes \cite{intro_shallow_water, weather, chanson, RAMMS}. 

The most commonly used model is the shallow water equations (SWE). This model can be derived from the Navier-Stokes equations by depth-averaging, assuming the horizontal velocity is the same throughout the height domain, i.e., the vertical velocity profile is constant. This greatly simplifies the system and decreases the computational cost. However, problems arise when the vertical velocity profile is varying throughout the height domain. In this case the model gives rise to large errors \cite{evaluating,intro_shallow_water}.

Recently a new model called the shallow water moment equations (SWME) has been developed \cite{intro_shallow_water}. The model assumes the vertical velocity profile to be a polynomial of a certain degree $N \in \mathbb{N}$. This increases the accuracy of the model while still maintaining some simplifying features. However, the polynomial velocity profile adds complexity in the form of additional variables (the polynomial coefficients). This means that the SWME requires more runtime than the more simple SWE.

To obtain a numerical solution, a PDE (e.g., SWME or SWE) is approximated by discretising the equations in space to obtain an ordinary differential equation (ODE). The resulting ODE is then discretised in time. This is often done using the Forward Euler method. Forward Euler is an explicit method, which means each time step is much cheaper than for implicit methods. However, it is not unconditionally stable so there is an upper bound on the time step size. If this upper bound is small then the numerical solution might demand prohibitively many time steps.

For that reason we are interested in the micro-macro method. The micro-macro is used to speed up the computation while retaining a sufficient level of accuracy of the solution. This is achieved by switching between two models; a high-order accurate micro model and a low-order less accurate macro model \cite{intro_micro_macro, monte_carlo}. However, so far the micro-macro method was only used for moment models of rarefied gases in \cite{intro_micro_macro} and not for free-surface flows modelled by the SWME \cite{intro_shallow_water}. 

The goal of this paper is to formulate the micro-macro method for the SWME to speed up free-surface flow simulations. The method will then be compared with the Forward Euler discretisation of the SWME in terms of accuracy and computational speed-up.

The remainder of this paper is organized as follows: In the rest of \cref{sec:1} we introduce the SWME and SWE models and the micro-macro method. In \cref{sec:2} we formulate the micro-macro method for SWME and provide a theoretical analysis. In \cref{sec:3} we provide numerical results from testing of the micro-macro method for SWME. 

\subsection{Moment Approximations for Shallow Flow Models}

\begin{figure}[H]
    \centering
    \includegraphics[width=0.9\linewidth]{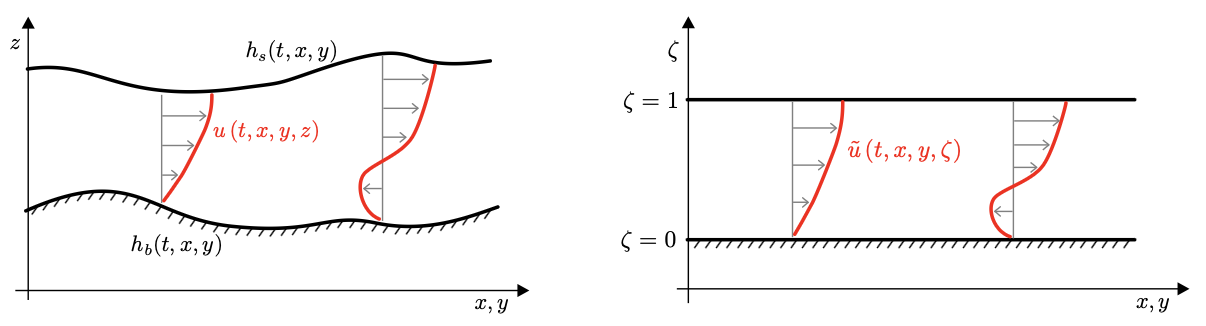}
    \caption{The vertical variable is mapped from irregular (left) to uniform domain (right), from \cite{intro_shallow_water}.}
    \label{cross_section}
\end{figure}

In this section we briefly recap the SWME, as introduced in \cite{intro_shallow_water}. We begin with the incompressible Navier-Stokes equations, which describe the velocity $u(t,x,y,z)$ of a fluid in space $(x,y,z)$ and time $t$. The fluid is bounded by the basal topography $h_b(t,x,y)$ and the free surface $h_s(t,x,y)$. We consider only a two dimensional problem in space and ignore the $y$ direction. To simplify the problem, the equations assume hydrostatic pressure, shallow flow, Newtonian fluid, and slip boundary condition. The system is scaled using the variable
\begin{equation*}
    \zeta(t,x,z) = \frac{z - h_b(t,x)}{h(t,x)},
\end{equation*}
where $h(t,x) = h_s(t,x)-h_b(t,x)$ is the fluid height. The variable $\zeta$ transforms the height domain of the equations to the interval $[0,1]$ as seen in \cref{cross_section} from \cite{intro_shallow_water}.

We assume the vertical velocity $u$ has a polynomial expansion of the form
\begin{equation}
    u(t, x,\zeta) = u_m(t, x) + \sum_{j=1}^N \alpha_j(t,x) \phi_j(\zeta),
    \label{u_expansion}
\end{equation}
where $u_m$ is the mean velocity of the flow, $\alpha_j$ are real coefficients and $\phi_j$ are basis functions. The functions $\phi_j$ are chosen to be the shifted Legendre polynomials on the interval $[0,1]$, normalised by $\phi_j(0) = 1$ \cite{intro_shallow_water}. These polynomials have many useful properties, including orthogonality with respect to the $L^2$ norm. This allows for simplification when deriving the evolution equations for the coefficients $\alpha_j$.

By depth-integrating the scaled equations against polynomial test functions and substituting expression \cref{u_expansion}, we obtain the SWME consisting of three parts \cite{intro_shallow_water}: First, mass balance equation is given by
\begin{equation}
    \pt h + \px (hu_m) = 0.
    \label{mass_balance}
\end{equation}

Second, the momentum balance equation may be written as
\begin{equation}
    \pt (hu_m) + \px \left( h \left(u_m^2 + \sum_{j=1}^N  \frac{\alpha_j^2}{2j+1}\right) + \frac{g}{2} e_z h^2 \right) = -\frac{\nu}{\lambda} \left( u_m + \sum_{j=1}^N \alpha_j \right) + hg(e_x-e_z \px h_b),
    \label{momentum_balance}
\end{equation}
where $\lambda$ is the bottom slip length and $\nu$ is the kinematic viscosity of the fluid. The gravitational acceleration $\textbf{g}$ is decomposed into its two directional components $\textbf{g} = g (e_x, e_z)$. In the case when the direction of the gravitational acceleration is aligned with the z-axis, then $e_x = 0$, $e_z = 1$.

Third, additional equations for $\alpha_i$ need to be derived to model the evolution of a non-constant velocity profile. 
These equations are derived from multiplying a reference equation with the Legendre polynomials $\phi_i$ for $i= 1, \ldots, N$. For more details we refer to \cite{intro_shallow_water}. The resulting equations are given by 
\begin{multline}
    \pt (h \alpha_i) + \px \left( h \left( 2u_m \alpha_i + \sum_{j,k=1}^N A_{ijk} \alpha_j \alpha_k \right) \right) \\
    = u_m \px (h\alpha_i) - \sum_{j,k=1}^N B_{ijk} \px (h\alpha_j) \alpha_k - (2i+1) \frac{\nu}{\lambda} \left( u_m + \sum_{j=1}^N \left( 1 \frac{\lambda}{h}C_{ij} \right) \alpha_j \right), 
    \label{additional_equations}
\end{multline}
where the constant values $A_{ijk}, B_{ijk}, C_{ij}$ are given by
\begin{equation*}
    A_{ijk} = (2i+1) \int_0^1 \phi_i \phi_j \phi_k \,d\zeta, \quad
    B_{ijk} = (2i+1) \int_0^1 \phi_i' \left( \int_0^\zeta \phi_j \,d\xi \right) \phi_k \,d\zeta, \quad
    C_{ij} = \int_0^1 \phi_i' \phi_j' \,d\zeta.
\end{equation*}
For any fixed value of $N$ the model consists of $N+2$ equations. It may be written in standard form as
\begin{equation}
    \pt U + A(U) \px U = S(U).
    \label{conservative_form}
\end{equation}
This is known as the non-conservative form. The term $A(U) \in \R^{(N+2)\times (N+2)}$ is called the system matrix and $S(U) \in \R^{N+2}$ is called the source term. It follows from the time derivative terms that $U = (h, hu_m, h\alpha_{1}, \ldots, h\alpha_{N})^T$.

In the rest of the paper we will assume that we have flat surface ($h_b$ is constant) and that gravity acts directly downwards ($e_x = 0$, $e_z = 1$).\\

As an example we consider the case when $N=0$. Then the system \eqref{conservative_form} simplifies to the SWE. In the non-conservative form they are given by
\begin{equation*}
    \pt 
    \begin{pmatrix}
        h \\ hu_m
    \end{pmatrix} + 
    \begin{pmatrix}
        0 & 1 \\ 
        gh - u_m^2 & 2u_m
    \end{pmatrix} \px
    \begin{pmatrix}
        h \\ hu_m
    \end{pmatrix} =
    \begin{pmatrix}
        0 \\ -\frac{\nu}{\lambda}u_m
    \end{pmatrix}.
\end{equation*}

It is evident from comparing the SWE and SWME that the SWE model is computationally less demanding because it contains only 2 equations, while the SWME model contains $N+2$ variables, leading to longer runtime and possibly smaller time steps.

\subsection{Micro-Macro Method}

The micro-macro method is a time discretisation method. It uses an accurate micro model of $M$ variables and a less accurate macro model of $L$ variables, such that $M>L$. The method switches between these two models to accelerate otherwise slow computations of the micro model. Compared to classical time discretization methods, the micro-macro method reduces runtime by using a simpler model for part of its inner steps. The used macro model has smaller size and therefore is less computationally expensive to work with. However, it costs runtime to switch between the models. For the micro-macro method to be useful, runtime reduction needs to be greater than the additional runtime costs.

\begin{figure}[H]
    \centering
    \includegraphics[width=0.6\linewidth]{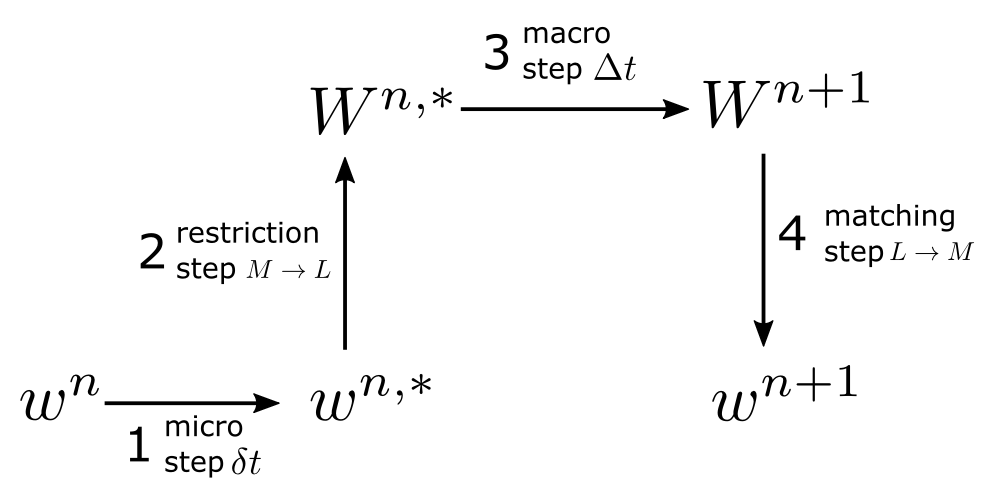}
    \caption{One iteration of the micro-macro method consists of micro step, restriction step, macro step, matching step, from \cite{intro_micro_macro}.}
    \label{mM}
\end{figure}

One iteration of the micro-macro method consists of four steps, see \cref{mM} from \cite{intro_micro_macro}:
\begin{enumerate}
    \item Simulation of the micro model with one time step.
    \item Restriction of the micro vector to the macro vector.
    \item Simulation of the macro model with one time step.
    \item Matching to reconstruct the micro vector from the macro vector.
\end{enumerate}

Let $w^n \in \R^M$ be the micro solution vector obtained after $n$ iterations of the micro-macro method. During the microscopic step we simulate one time step of the micro model with step size $\delta t$. We obtain an intermediate micro solution $w^{n,*} \in \R^M$.

In the restriction step we restrict $w^{n,*}$ to the first $L$ variables to obtain an intermediate macro solution $W^{n,*} \in \R^L$.

In the macroscopic step we simulate one time step of the macro model with step size $\Delta t$ to obtain a macro solution $W^{n+1} \in \R^L$.

During the matching step we reconstruct the micro solution $w^{n+1} \in \R^M$ from the macro solution $W^{n+1}$. However the dimension of $W^{n+1}$ is less than that of $w^{n+1}$, and $W^{n+1}$ therefore does not provide all the necessary information. The information is filled in by using the last available micro solution vector $w^{n,*}$. We do this by minimising the distance $d(w^{n,*}, w^{n+1})$ with respect to some problem-specific pseudo norm $d(\cdot,\cdot)$ \cite{intro_micro_macro, monte_carlo}. 

If an explicit discretisation method is chosen for both the microscopic step (1.) and macroscopic step (3.), the time steps $\delta t$ and $\Delta t$ have to be chosen such that they ensure stability of the method. Typically the bound on the microscopic step is smaller than on the macroscopic step, meaning that $\Delta t > \delta t$. This means that the micro-macro method reduces runtime with respect to a pure micro model simulation not just by using a smaller macro model during the macro steps, but also by potentially using larger macro time steps.

\section{Formulating the Micro-Macro Method for Shallow Flow Models}
\label{sec:2}

In this section we formulate the four steps of the micro-macro method for SWME. The micro and macro models are defined by \cref{mass_balance}, \cref{momentum_balance} and \cref{additional_equations} via particular choices of the number of equations such that the micro model uses $M$ variables and the macro model uses $L$ variables, with $M >L$. 

\textbf{Microscopic step} Discretise the micro model in time using Forward Euler and perform one step with time step $\delta t$. This is done by solving the equation
\begin{equation*}
    \begin{pmatrix}
        h \\ hu_m \\ h\alpha_1 \\ \vdots \\ h\alpha_{M-2}
    \end{pmatrix}^{n,*} = 
    \begin{pmatrix}
        h \\ hu_m \\ h\alpha_1 \\ \vdots \\ h\alpha_{M-2}
    \end{pmatrix}^{n} + \delta t \text{ }f
     \begin{pmatrix}
        h \\ hu_m \\ h\alpha_1 \\ \vdots \\ h\alpha_{M-2}
    \end{pmatrix}^{n},
\end{equation*}
where $f(h, u_m, \alpha_1, \ldots, \alpha_{M-2})^n$ is a function including the spatial discretization of the system matrix term and the source term of the micro model, evaluated at the time corresponding to $w^n$.

\textbf{Restriction} The vertical velocity profile $u^{n,*}$, see \eqref{u_expansion}, corresponding to the variable vector $w^{n,*} \in \R^{M}$, is a polynomial of degree $M-2$. We wish to restrict $w^{n,*}$ such that the resulting polynomial is of degree $L-2$. Since the expansion \eqref{u_expansion} uses an orthogonal basis, we do this by omitting the coefficients in \cref{u_expansion} corresponding to the higher degree polynomials. Then we get the restricted micro solution, which will be the starting point for the macro step.
\begin{equation*}
    W^{n,*} = \begin{pmatrix}
        h \\ hu_m \\ h\alpha_1 \\ \vdots \\ h\alpha_{L-2}
    \end{pmatrix}^{n,*}.
\end{equation*}
\textbf{Macroscopic step} Discretise the macro model in time using Forward Euler and perform one step with time step $\Delta t$ by solving the equation
\begin{equation*}
    \begin{pmatrix}
        h \\ hu_m \\ h\alpha_1 \\ \vdots \\ h\alpha_{L-2}
    \end{pmatrix}^{n+1} = 
    \begin{pmatrix}
        h \\ hu_m \\ h\alpha_1 \\ \vdots \\ h\alpha_{L-2}
    \end{pmatrix}^{n,*} + \Delta t \text{ }F
     \begin{pmatrix}
        h \\ hu_m \\ h\alpha_1 \\ \vdots \\ h\alpha_{L-2}
    \end{pmatrix}^{n,*} .
\end{equation*}
Here $F(h, u_m, \alpha_1, \ldots, \alpha_{L-2})^{n,*}$ is a function including the  spatial discretization of the system matrix term and the source term of the macro model, evaluated at the time corresponding to $W^{n,*}$.

\textbf{Matching} Due to the orthogonality of the expansion \eqref{u_expansion}, we can match the first $L$ elements of $w^{n+1}$ by setting them equal to $W^{n+1}$. For the remaining $M-L$ coefficients $\alpha_i$ we minimise the distance between $w^{n,*}$ and $ w^{n+1}$ with respect to some pseudo norm. Denote the velocity profiles corresponding to $w^{n,*}$ and $ w^{n+1}$ as $u^{n,*}$ and $u^{n+1}$, respectively. We choose to minimise the distance $d(u^{n+1}, u^{n,*})$ with respect to the $L^2$ norm. Using the $L^2$ norm allows us to take advantage of the properties of the Legendre polynomials. 

Then
\begin{equation*}
    u^{n+1} = \argmin_{\bar{u} \in V(W^{n+1})} d(\bar{u}, u^{n,*}),
\end{equation*}
where $V(W^{n+1})$ is the set of all velocity profiles of the form \cref{u_expansion} constrained by the first variables $W^{n+1}$. This equation can be solved explicitly. The distance $d(\bar{u}, u^{n,*})$ is given by
\begin{align*}
    d(\bar{u}, u^{n,*}) &= \int_0^1 \left(\bar{u}(\zeta)- u^{n,*}(\zeta) \right)^2 \,d\zeta \\
    &= \int_0^1 \left( \bar{u}_m+\sum^N_{j=1} \bar{\alpha}_j\phi_j(\zeta) - u_m - \sum^N_{j=1}\alpha_j \phi_j(\zeta) \right)^2 \,d\zeta \\
    &= \int_0^1 \left( (\bar{u}_m - u_m) + \sum^N_{j=1}(\bar{\alpha}_j -\alpha_j) \phi_j(\zeta) \right)^2 \,d\zeta \\
    &= \int_0^1 (\bar{u}_m - u_m)^2 + 2(\bar{u}_m - u_m) \sum^N_{j=1}(\bar{\alpha}_j -\alpha_j) \phi_j(\zeta) + \left( \sum^N_{j=1}(\bar{\alpha}_j -\alpha_j) \phi_j(\zeta) \right)^2 \,d\zeta.
\end{align*}

Here we take advantage of two properties of the shifted Legendre polynomials. Let $\phi_i,\phi_j$ denote the i-th and j-th shifted Legendre polynomial on $[0,1]$, respectively. Then
\begin{equation*}
    \int_0^1 \phi_i (\zeta) \,d\zeta = 0, \textrm{ and }
    \int_0^1 \phi_i(\zeta) \phi_j (\zeta) \,d\zeta = 0
\end{equation*}
for all $i,j \geq 1$ such that $i\neq j$. Using the first property, the second term of the integral drops out. Using the second property, the third term of the integral may be simplified. All cross terms of the product drop out, leaving only the squared terms. 
We are left with
\begin{align*}
    d(\bar{u}, u^{n,*}) &= \int_0^1 (\bar{u}_m - u_m)^2 \,d\zeta + \int_0^1 \sum^N_{j=1}(\bar{\alpha}_j -\alpha_j)^2 \phi_j^2(\zeta) \,d\zeta \\
    &= \int_0^1 (\bar{u}_m - u_m)^2 \,d\zeta + \sum^N_{j=1}(\bar{\alpha}_j -\alpha_j)^2 \int_0^1 \phi_j^2(\zeta) \,d\zeta.
\end{align*}

The distance is minimal if the derivative with respect to $\alpha_{L-1}, \ldots, \alpha_{M-2}$ is zero. This derivative is given by
\begin{align*}
    \frac{\partial d(\bar{u}, u^{n,*})}{\partial \bar{\alpha}_i} &= 2(\bar{\alpha}_i -\alpha_i) \int_0^1 \phi_i^2(\zeta) \,d\zeta.
\end{align*}

By setting the derivative above to be equal to zero, it follows immediately that $\bar{\alpha}_i =\alpha_i$. This means that the last $M-L$ coefficients $\alpha_i$ are simply carried over from $u^{n,*}$ to $u^{n+1}$. We use this to calculate the remaining elements of $w^{n+1}$. Note that this crucially depends on the orthogonality of the Legendre basis functions with respect to the $L^2$ norm. For other basis functions or norms, the matching step might be much more involved.

The time step sizes $\delta t$ and $\Delta t$ require further attention. As the micro-macro method uses Forward Euler time discretisation within both the micro step and the macro step, there is a bound on the step size to ensure stability of each step. For non-conservative equations of the form $\pt U + A(U) \px U = 0$ the bound is given by the Courant–Friedrichs–Lewy (CFL) condition, which reads 
\begin{equation*}
    \frac{\Delta t}{\Delta x} \leq \frac{C}{|\lambda|_{max}},
\end{equation*}
where $|\lambda|_{max}$ is the magnitude of the largest eigenvalue of $A(U)$ and $C$ is a constant such that $C \leq 1$ \cite{CFL}. We choose $\delta t$ and $\Delta t$ to be the maximum step sizes allowed.

Note that the CFL condition above does not consider the source term of our equations. This may lead to stability issues in some test cases. This issue is further discussed in section \ref{sec:time_step_analysis}.

For the SWME derived in \cite{intro_shallow_water}, a closed form of $|\lambda|_{max}$ exists only for $A(U)$ of dimension $2\times 2$ or $3\times 3$, and computing the exact eigenvalue for every $A(U)$ would be costly. Therefore we use the approximation
\begin{equation}
    |\lambda|_{max} = |u_m| + \sqrt{gh + \sum_{j=1}^N \alpha_j^2}.
    \label{eigenvalue}
\end{equation}
This expression is exact for $N = 0$ and $N=1$. For higher values of $N$, it gives slightly higher values than the exact eigenvalues, which ensures the stability condition is not violated. A similar approach is used in the numerical computations in \cite{intro_shallow_water}. 

The micro-macro method for SWME is implemented in Python. We use a first-order finite volume space discretisation. Additionally, both the micro and macro steps use a splitting scheme. We refer to the GitHub repository of our code  for more details \cite{github}.


\subsection{Investigation of the Matching}

During the restriction step, information about the micro variables is lost. We want to investigate how much of the information is recovered during the matching step. To test this we consider $w^{n,*}$ to be a typical velocity profile with $M=7$ (i.e., $N=M-2=5$) given by example values 
\begin{equation*}
    h = 1.406, u_m=0.352, \alpha_1= -0.120, \alpha_2= -0.096, \alpha_3= -0.042, \alpha_4= -0.006, \alpha_5= 0.003.
\end{equation*}

\begin{figure}[H]
    \centering
    \includegraphics[width=0.48\linewidth]{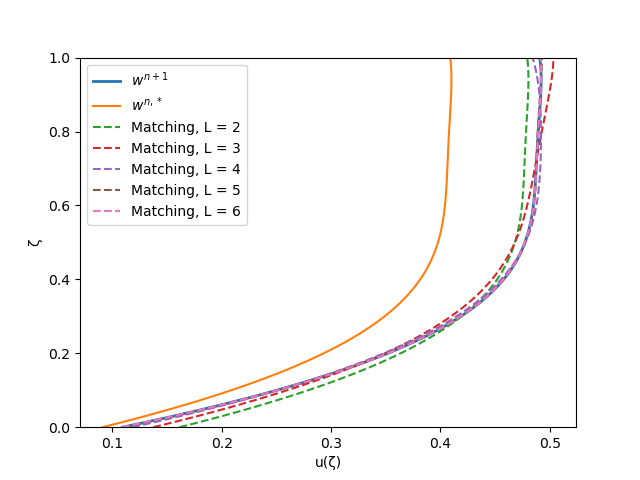}
    \includegraphics[width=0.48\linewidth]{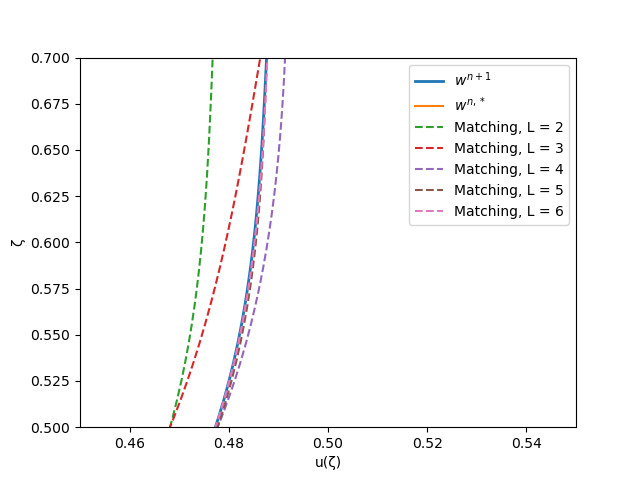}
    \caption{Investigation of the matching step for $L = 2, \ldots, 6$ variables. Resulting velocity profiles (left) and detailed view of $\zeta \in [0.5, 0.7]$ (right). Increasing the number of variables leads to a more accurate matching.}
    \label{matching}
\end{figure}

We also assume the exact micro solution $w^{n+1}$ is such that the variables $h, u_m, \alpha_1, \ldots, \alpha_5$ are scaled by a factor of $1.2$. We perform the matching step for $L = 2, \ldots, 6$ macro variables as described in the previous section. Note that for $L=7$, the matching produces the exact solution. The results are shown in \cref{matching}.

The left figure of \cref{matching} shows a comparison of the velocity profiles across the entire domain. We see that the matching procedure generates quite accurate results even when only 2 variables are taken directly. The right figure is a zoomed in view to $[0.5, 0.7]$ which shows that the matched velocity profiles converge to the exact micro solution. Increasing the number of variables taken directly leads to a more accurate matching. We conclude that the matching step works well and does not lead to large errors. Note that similar results were obtained in \cite{intro_micro_macro} for a gas dynamic moment model.

\subsection{Computational Complexity}

The main goal of the micro-macro method is to speed up the computation time compared to classical methods. In this section we calculate the computational complexity of the micro-macro method for shallow flows and compare it to a classical microscopic model. The complexity clearly grows linearly with the number of grid cells for both methods, therefore we consider the complexity per grid cell, same as in \cite{intro_micro_macro}.

The classical microscopic model (without micro-macro) uses $M$ variables. The system matrix of the model is of size $M \times M$. Then for one time step the complexity is of order $\mathcal{O}(M^2)$ per grid cell. The largest possible time step $\delta t$ while still satisfying the CFL condition is given by $\delta t = \frac{\Delta x}{|\lambda|_{max, M}}C$, which is of order $\mathcal{O}(\Delta x)$. Then the computational complexity per grid cell is
\begin{equation}
    \mathcal{O} \left(\frac{M^2}{\Delta x} \right).
    \label{o_1}
\end{equation}

The computational complexity of the micro-macro method is determined by the sum of complexities of its four steps. It is given as follows:
\begin{enumerate}
    \item The micro model uses $M$ variables so the system matrix is of size $M \times M$. The complexity of the micro step is $\mathcal{O}(M^2)$ per grid cell and time step.
    \item The restriction step is a linear mapping of $L$ variables therefore the complexity is $\mathcal{O}(L)$ per grid cell and time step.
    \item The macro model uses $L$ variables so the system matrix is of size $L \times L$. The complexity of the macro step is $\mathcal{O}(L^2)$ per grid cell and time step.
    \item The matching step is a linear mapping of $M-L$ variables from the intermediate micro solution. The complexity is $\mathcal{O}(M-L)$ per grid cell and time step.
\end{enumerate}

Then the complexity of one iteration of the micro-macro method is 
\begin{equation*}
    \mathcal{O} (M^2+L^2 + M).
\end{equation*}
per grid cell. The length of one full iteration of the micro-macro method is equivalent to $\delta t + \Delta t$. The step sizes $\delta t $ and $ \Delta t$ are determined by the CFL condition such that $\delta t = \Delta x \frac{C}{|\lambda|_{max, M}}$, $\Delta t = \Delta x\frac{C}{|\lambda|_{max, L}}$. For any fixed end time of the algorithm, the values $|\lambda|_{max, M}$ and $|\lambda|_{max, L}$ may be bounded by a constant, therefore both of the time step sizes $\delta t $ and $ \Delta t$ are considered $\mathcal{O}(\Delta x)$. Then the complexity per grid cell of the micro-macro method is
    

\begin{equation}
    \mathcal{O} \left(\frac{M^2+L^2 + M}{2 \Delta x} \right).
    \label{o_2}
\end{equation}

The calculations done here do not allow us to make an accurate comparison of the two methods, since most of the constants have been omitted. However, we may deduce from \cref{o_1} and \cref{o_2} that if $M$ is much larger than $L$, the micro-macro method is likely to be faster than the micro model. Note that this result is similar to the result in \cite{intro_micro_macro} for moment models of gas dynamics.

\subsection{Analysis of the Time Step}
\label{sec:time_step_analysis}

The implementation of the micro-macro method uses a splitting scheme for the micro and macro steps. This splitting scheme separates the differential equation \cref{conservative_form} into the equations

\begin{align}
    \pt U + A(U) \px U  = & ~0,    \label{p1} \\
    \pt U   \hspace{2.05cm}             = & ~S(U). \label{p2}
\end{align}
The micro and macro steps each perform one time step of the equations above.

The equations \cref{p1} and \cref{p2} are solved explicitly and therefore require a bound on the step size to ensure stability of the method. However, the implementation only uses the CFL condition to determine the step size. This ensures stability for \cref{p1}, but not for \cref{p2}. This limits the usage of the model.


Denote the bound on the time step size for \cref{p1} as $\Delta t_1$ and for \cref{p2} as $\Delta t_2$. In the case when $\Delta t_1 > \Delta t_2$, the stability of the method is not guaranteed. This may occur in the two following situations:
\begin{enumerate}
    \item The source term is much larger than zero.
    \item The bound $\Delta t_1$ is very large.
\end{enumerate}

The source term of the shallow water moment equations includes the friction terms. Notice that those terms in equations \cref{mass_balance}, \cref{momentum_balance} and \cref{additional_equations} are all multiplied by the factor $\frac{\nu}{\lambda}$. Therefore the source term is large if the magnitude of $\frac{\nu}{\lambda}$ is very large.

The bound $\Delta t_1$ is large in two cases. Firstly, if the grid cell size $\Delta x$ is chosen to be large. Second, if $|\lambda|_{max}$ is close to zero. This happens when the variables $h, u_m, \alpha_1, \ldots, \alpha_N$ are all close to zero.

The specific implementation of the micro-macro method should not be used in cases where $\left| \frac{\nu}{\lambda} \right|$ is large, $\Delta x$ is large or $h, u_m, \alpha_1, \ldots, \alpha_N$ are all close to zero. In those cases, the method is not necessarily stable.

We note that there exist ways to overcome stability problems for stiff right hand side terms, as outlined in \cite{KoellermeierQian2022}.


\section{Results}
\label{sec:3}

We perform a series of tests to determine the accuracy of the solution and computational speed of the micro-macro method. We present results for two different test cases, in which the micro-macro method is tested for varying numbers of micro and macro variables and compared to a solution of a macro model of $L=2$ variables and a micro model of $M=7$ variables. An exact solution of the Navier-Stokes equations does not exist so we treat the micro solution (i.e., with $M=7$) as a reference.

\subsection{Dam Break Test}

We first consider a dam break test case similar to the one also used in \cite{intro_shallow_water}. The initial flow velocity is zero and the height profile has a discontinuity at $x=0$. The initial condition is thus given by
\begin{equation*}
    w(0,x) = 
    \begin{cases}
        (2,0, \ldots, 0)^T, & \text{if } x \leq 0 \\
        (1, 0, \ldots, 0)^T, & \text{if } x > 0.
    \end{cases}
\end{equation*}

For the simulation we compute the solution on the spatial domain $[-4,4]$ partitioned into 400 cells. The slip length and viscosity are fixed to be $\lambda = \nu = 0.1$, and the end time is $t = 2$. 

\begin{figure}[H]
    \centering
    \includegraphics[width=1\linewidth]{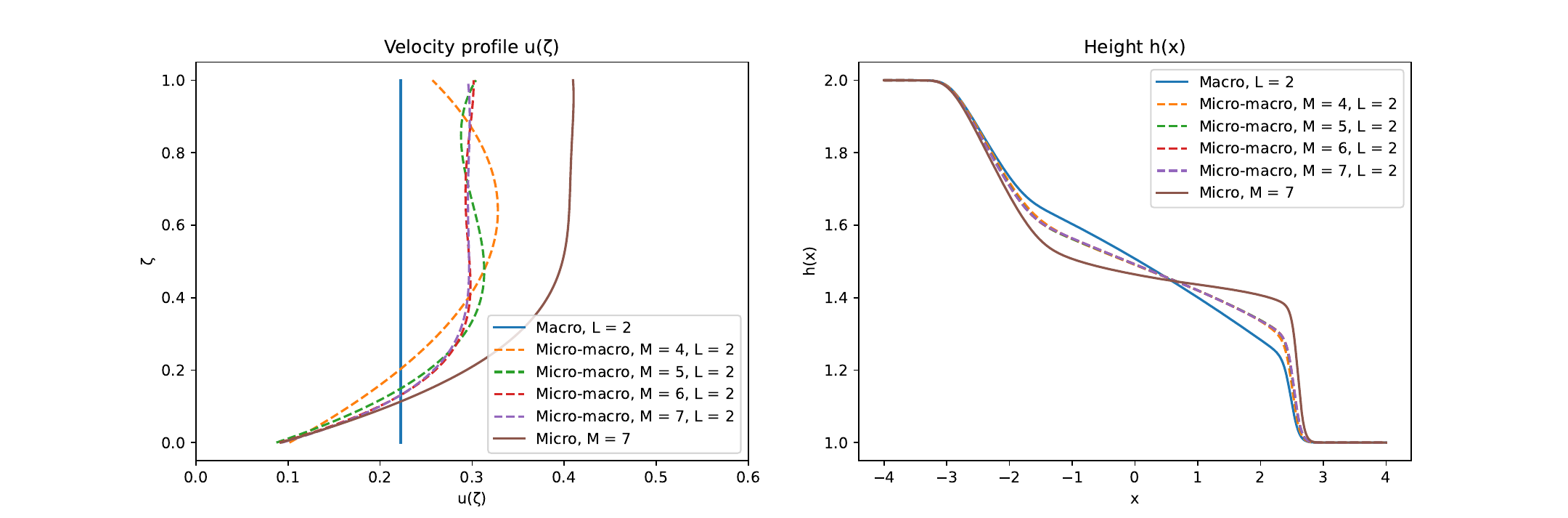}
    \caption{Micro-macro method simulation results for the dam break test. The macro model is fixed to be $L=2$ and the micro model is varied between $M= 4,\ldots, 7$. Velocity profile $u(\zeta)$ (left) and water height $h(x)$ (right).}
    \label{db0}
\end{figure}

\begin{figure}[H]
    \centering
    \includegraphics[width=1\linewidth]{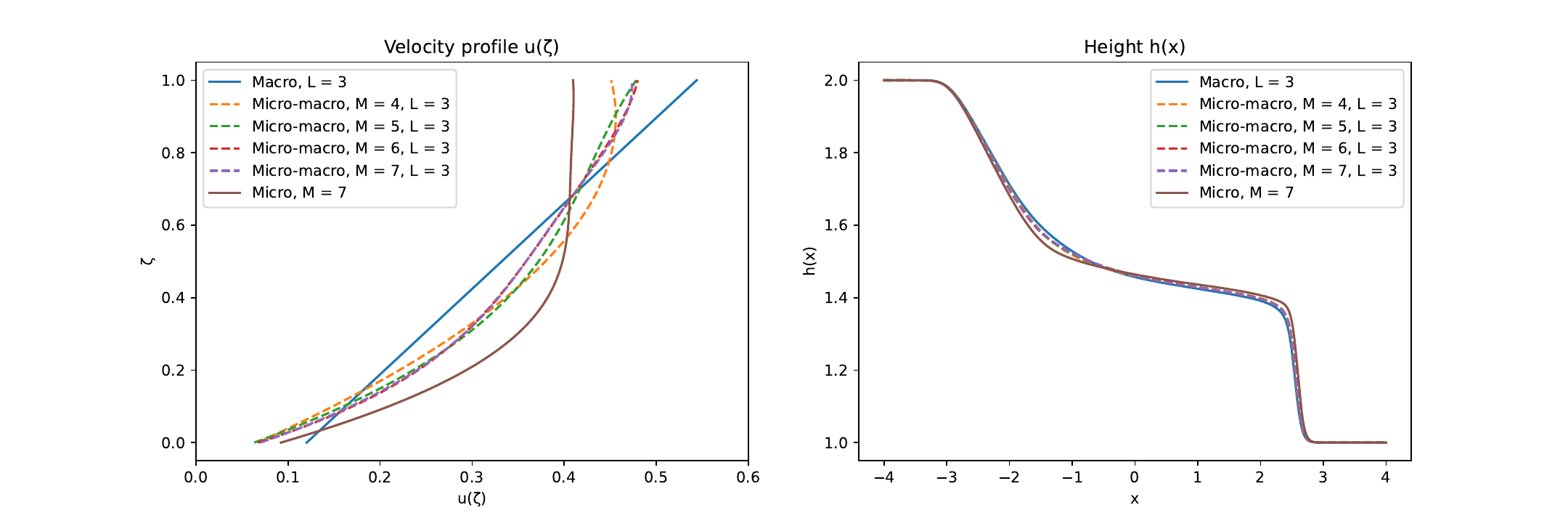}
    \caption{Micro-macro method simulation results for the dam break test. The macro model is fixed to be $L=3$ and the micro model is varied between $M = 4,\ldots, 7$.  Velocity profile $u(\zeta)$ (left) and water height $h(x)$ (right).}
    \label{db1}
\end{figure}

In \cref{db0} the macro model is fixed to have $L=2$ variables and the micro model is varied between $M = 4,\ldots, 7$ variables. The left figure shows the velocity profile $u(\zeta)$ at $x=2$ and the right figure shows the water height $h(x)$. Increasing the number of micro variables leads to a more accurate velocity profile. However, the velocity profile does not converge to the reference solution. Similarly, the height function does not converge to the reference micro height.

In \cref{db1} the macro model is fixed to have $L= 3$ variables and the micro model is varied between $M = 4,\ldots, 7$ variables. The water height and velocity profiles are both much closer to the reference solution than in the previous case. However, as the number of micro variables increases, the velocity profile still does not converge to the reference solution.

\begin{figure}[H]
    \centering
    \includegraphics[width=1\linewidth]{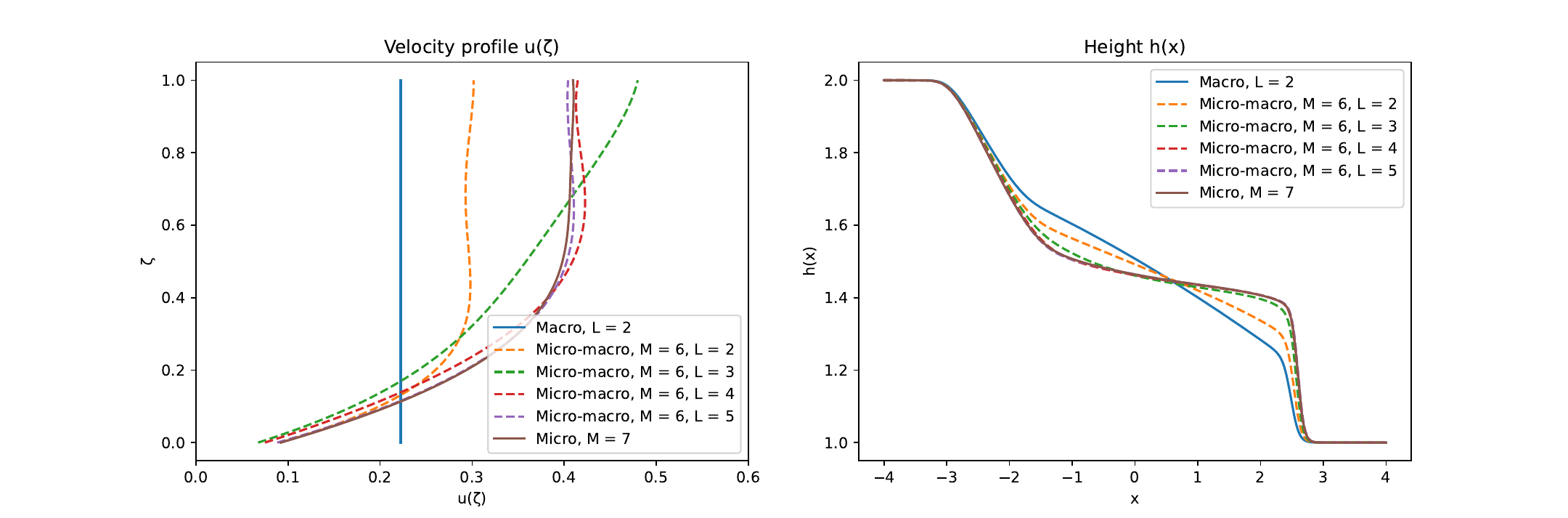}
    \caption{Micro-macro method simulation results for the dam break test. The micro model is fixed to be $M=6$ and the macro model is varied between $L = 2,\ldots, 5$.  Velocity profile $u(\zeta)$ (left) and water height $h(x)$ (right).
    }
    \label{db_fix_m_1}
\end{figure}

\begin{figure}[H]
    \centering
    \includegraphics[width=1\linewidth]{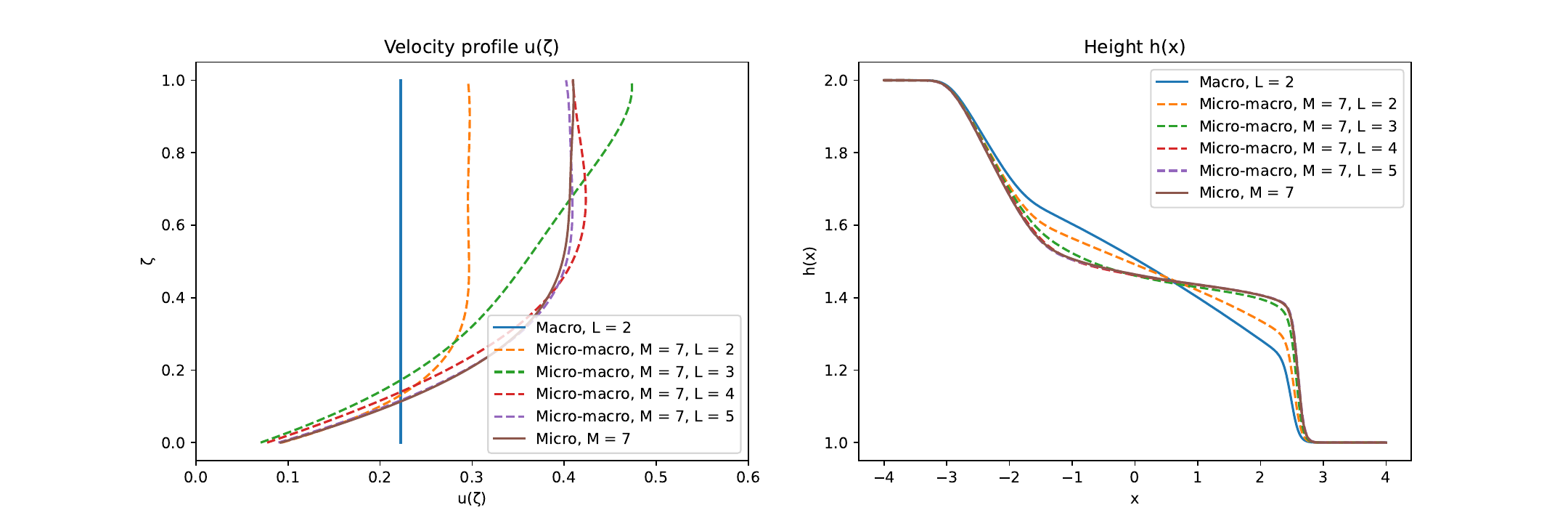}
    \caption{Micro-macro method simulation results for the dam break test. The micro model is fixed to be $M=7$ and the macro model is varied between $L = 2,\ldots, 5$.  Velocity profile $u(\zeta)$ (left) and water height $h(x)$ (right).}
    \label{db_fix_m_2}
\end{figure}

In \cref{db_fix_m_1} the micro model is fixed to be $M=6$ and the macro model is varied between $L = 2,\ldots, 5$. For lower values of $L$ the velocity profile is not accurate. But as $L$ increases, the velocity profile of the micro-macro method converges to the reference solution. The same is true for the water height.

In \cref{db_fix_m_2} the micro model is fixed to be $M=7$ and the macro model is varied for $L = 2,\ldots, 5$. The results shown are slightly more accurate than in the previous case, but the difference is barely noticeable. Increasing the number of macro variables $L$ leads to a convergence to the reference solution.

Next we compare the computational speed of the tested methods. \cref{tab1} and \cref{tab2} show the computational speed-up of the micro-macro method with respect to the micro model. We see that the micro-macro method is consistently faster for all parameters tested. The tables also show the speed-up for the macro model. As expected, the macro model is the fastest. However, this comes at the cost of some loss of accuracy as investigated above. 

\begin{table}[H]
    \centering
    \caption{Micro-macro method speed-up with respect to the reference micro solution for the dam break test varying $M$.}
    \begin{tabular}{|m{4cm}| m{1.2cm} m{1.2cm}|}
        \hline 
         & \multicolumn{2}{c|}{Speed-up} \\ \hline
        Number of variables & $L=2$ & $L=3$ \\ \hline
        Macro & 2.925 & 2,478 \\ \hline
        Micro-macro, $M=4$ & 2.583 & 2.387 \\ 
        Micro-macro, $M=5$ & 2.265 & 2.137 \\ 
        Micro-macro, $M=6$ & 1.924 & 1.831 \\ 
        Micro-macro, $M=7$ & 1.619 & 1.547 \\ \hline
        Micro, $M=7$ (ref.) & \multicolumn{2}{c|}{1} \\ \hline
    \end{tabular}
    \label{tab1}
\end{table}

\begin{table}[H]
    \centering
    \caption{Micro-macro method speed-up with respect to the reference micro solution for the dam break test varying $L$.}
    \begin{tabular}{|m{4cm}| m{1.2cm} m{1.2cm}|}
        \hline 
         & \multicolumn{2}{c|}{Speed-up} \\ \hline
        Macro, $L=2$& \multicolumn{2}{c|}{2.925} \\ \hline
        Number of variables & $M=6$ & $M=7$ \\ \hline
        Micro-macro, $L=2$ & 1.924 & 1.619 \\ 
        Micro-macro, $L=3$ & 1.831 & 1.547 \\ 
        Micro-macro, $L=4$ & 1.706 & 1.463 \\ 
        Micro-macro, $L=5$ & 1.567 & 1.342 \\ \hline
        Micro, $M=7$ (ref.) & \multicolumn{2}{c|}{1} \\ \hline
    \end{tabular}
    \label{tab2}
\end{table}

In summary, the micro-macro method succeeds at obtaining accurate but fast solutions for the dam-break test case with speed-ups of more than a factor of two.

\subsection{Wave Transport Test}
We consider a second test case which represents the transport of a wave. It is given by the initial condition
\begin{equation*}
    w(0,x) = (3+ e^{-1.5x^2},0, \ldots, 0)^T. 
\end{equation*}
The solution is calculated on the spatial domain $[-4,4]$ partitioned into 400 cells. The slip length and viscosity are fixed to be $\lambda = \nu = 0.1$, and the end time is $t = 1$.

\begin{figure}[H]
    \centering
    \includegraphics[width=\linewidth]{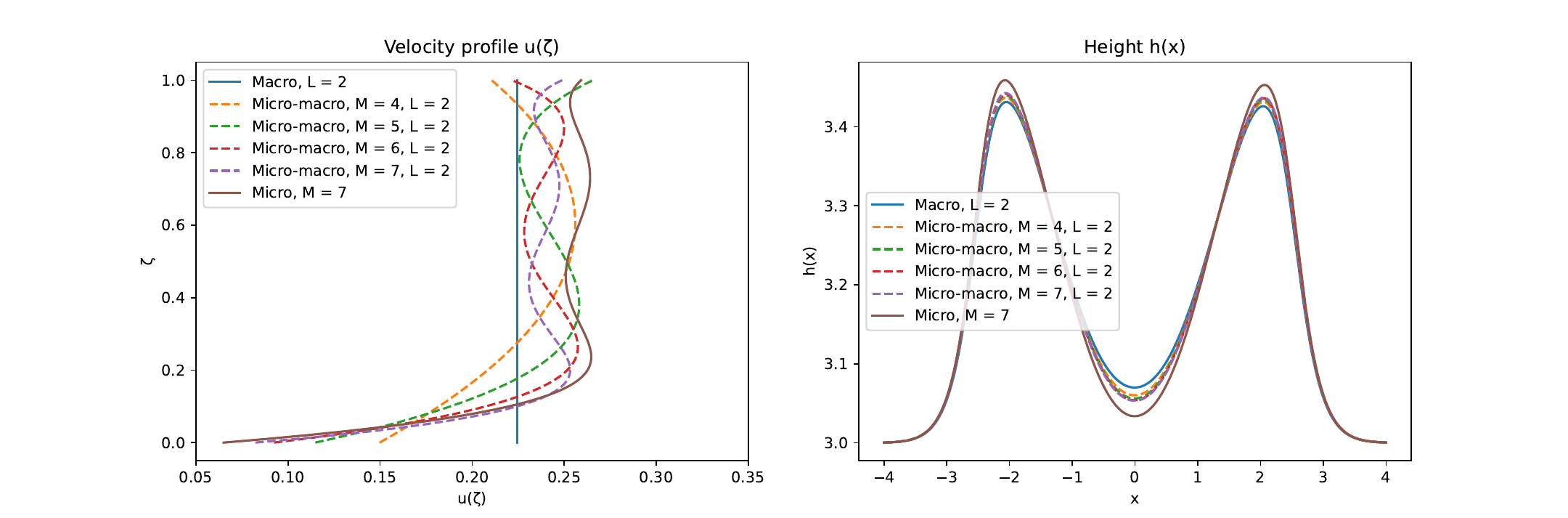}
    \caption{Micro-macro method simulation results for the wave transport test. The macro model is fixed to be $L=2$ and the micro model is varied between $M = 4,\ldots, 7$.  Velocity profile $u(\zeta)$ (left) and water height $h(x)$ (right).}
    \label{sm_fix_l_1}
\end{figure}

\begin{figure}[H]
    \centering
    \includegraphics[width=\linewidth]{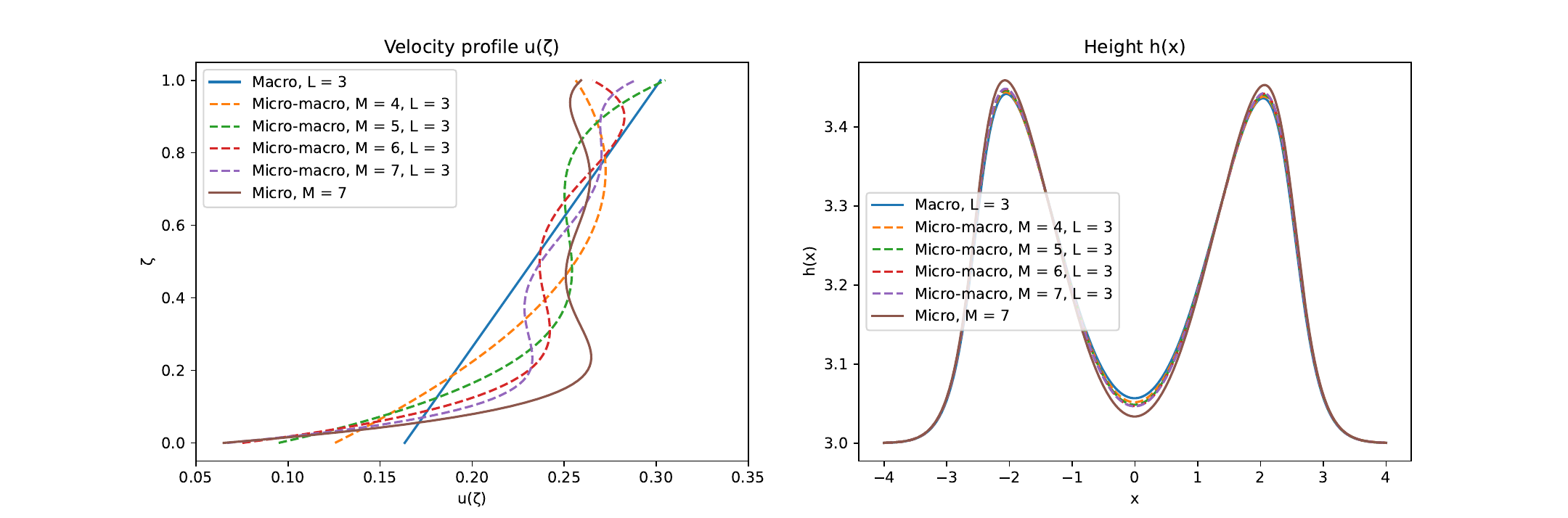}
    \caption{Micro-macro method simulation results for the wave transport test. The macro model is fixed to be $L=3$ and the micro model is varied between $M = 4,\ldots, 7$.  Velocity profile $u(\zeta)$ (left) and water height $h(x)$ (right).}
    \label{sm_fix_l_2}
\end{figure}

In \cref{sm_fix_l_1} the macro model is fixed to have $L=2$ variables and the micro model is varied between $M = 4,\ldots, 7$ variables. The left figure shows the velocity profile $u(\zeta)$ at $x=2$ and the right figure shows the height $h(x)$. All the velocity profiles of the micro-macro method are in between the micro model and the macro model. As $M$ increases the velocity profile is still changing with $M$. However, the water height seems to converge with $M$.

In \cref{sm_fix_l_2} the macro model is fixed to have $L=2$ variables and the micro model is varied between $M=4$ and $M=7$ variables. The height and velocity profiles are both slightly closer to the reference solution than in the previous case. We again do not observe any clear behaviour of the velocity profiles as $M$ increases.

\begin{figure}[H]
    \centering
    \includegraphics[width=\linewidth]{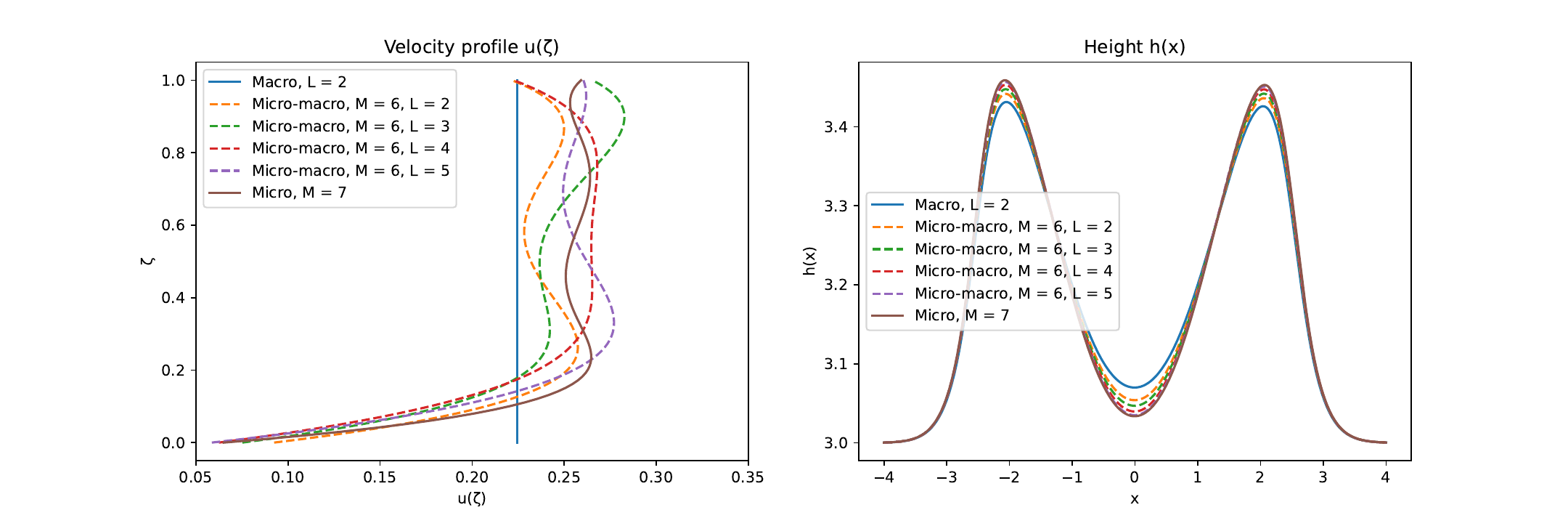}
    \caption{Micro-macro method simulation results for the wave transport test. The micro model is fixed to be $M=6$ and the macro model is varied between $L = 2,\ldots, 5$.  Velocity profile $u(\zeta)$ (left) and water height $h(x)$ (right).}
    \label{sm_fix_m_1}
\end{figure}

\begin{figure}[H]
    \centering
    \includegraphics[width=\linewidth]{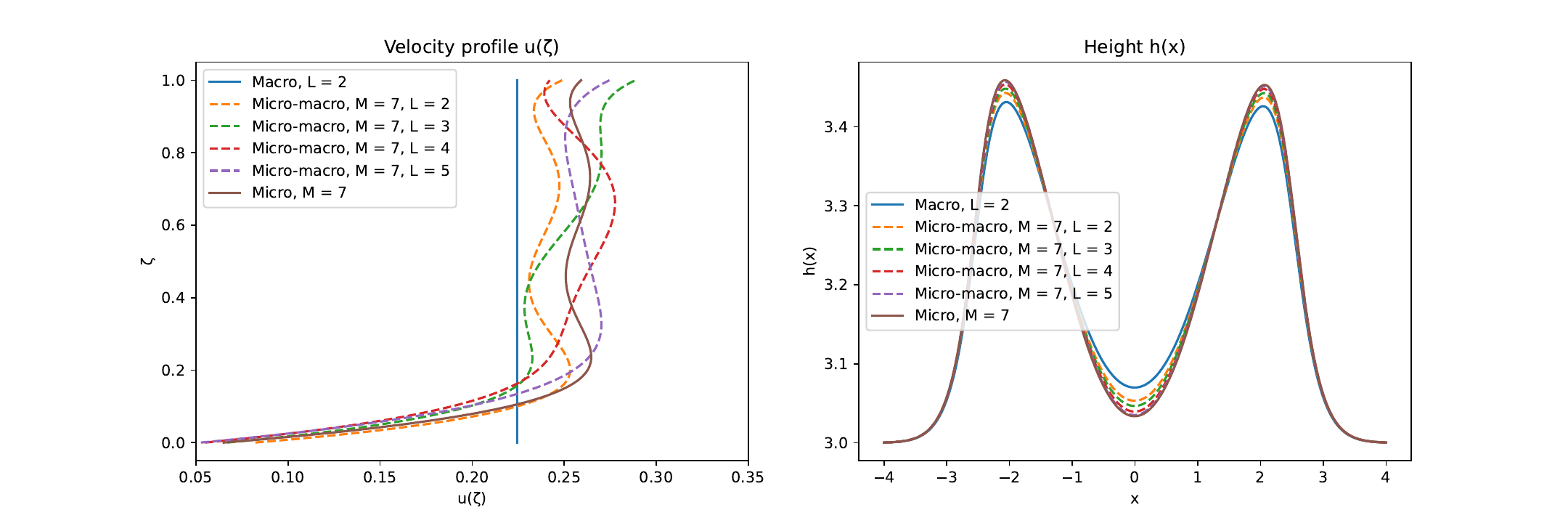}
    \caption{Micro-macro method simulation results for the wave transport test. The micro model is fixed to be $M=7$ and the macro model is varied between $L = 2,\ldots, 5$.  Velocity profile $u(\zeta)$ (left) and water height $h(x)$ (right).}
    \label{sm_fix_m_2}
\end{figure}

In \cref{sm_fix_m_1} the micro model is fixed to be $M=6$ and the macro model is varied for $L = 2,\ldots, 5$. As $L$ increases, the accuracy of the height and velocity profile significantly improves. However, the velocity profile does not seem to converge to the reference solution. On the other hand, the height of the micro-macro method does converge to the reference solution.

In \cref{sm_fix_m_2} the micro model is fixed to be $M=7$ and the macro model is varied for $L = 2,\ldots, 5$. The results shown are more accurate than in the previous case, especially for the velocity profile. We again observe convergence only for the water height.

\cref{tab3} and \cref{tab4} show the computational speed-up of the micro-macro method with respect to the reference solution. The micro-macro method is consistently faster for all values tested. An interesting observation is that for any fixed $M$ and $L$, the speed-up of the method is very similar to the speed-up shown in \cref{tab1} and \cref{tab2}. This indicates the computational complexity of the method does not depend on the initial conditions and the results found in this work are relatively general.

\begin{table}[H]
    \centering
    \caption{Micro-macro method speed-up w.r.t.\! the reference micro solution for the wave transport test varying $M$.}
    \begin{tabular}{|m{4cm}| m{1.2cm} m{1.2cm}|}
        \hline 
         & \multicolumn{2}{c|}{Speed-up} \\ \hline
        Number of variables & $L=2$ & $L=3$ \\ \hline
        Macro, $L=2$ & 2.833 & 2.467 \\ \hline
        Micro-macro, $M=4$ & 2.482 & 2.358 \\ 
        Micro-macro, $M=5$ & 2.202 & 2.103 \\ 
        Micro-macro, $M=6$ & 1.875 & 1.805 \\ 
        Micro-macro, $M=7$ & 1.574 & 1.521 \\ \hline
        Micro, $M=7$ (ref.) & \multicolumn{2}{c|}{1} \\ \hline
    \end{tabular}
    \label{tab3}
\end{table}

\begin{table}[H]
    \centering
    \caption{Micro-macro method speed-up w.r.t.\! the reference micro solution for the wave transport test varying $L$.}
    \begin{tabular}{|m{4cm}| m{1.2cm} m{1.2cm}|}
        \hline 
         & \multicolumn{2}{c|}{Speed-up} \\ \hline
        Macro, $L=2$& \multicolumn{2}{c|}{2.833} \\ \hline
        Number of variables & $M=6$ & $M=7$ \\ \hline
        Micro-macro, $L=2$ & 1.875 & 1.574 \\ 
        Micro-macro, $L=3$ & 1.805 & 1.521 \\ 
        Micro-macro, $L=4$ & 1.693 & 1.441 \\ 
        Micro-macro, $L=5$ & 1.561 & 1.244 \\ \hline
        Micro, $M=7$ (ref.) & \multicolumn{2}{c|}{1} \\ \hline
    \end{tabular}
    \label{tab4}
\end{table}

In summary, the micro-macro method computes increasingly accurate water heights with speed-ups of more than a factor of two also for the wave transport test.

\section{Conclusion}

We formulated a micro-macro method for shallow water moment equations with the aim to be more accurate than simple a macroscopic model and faster than accurate microscopic models. We derived the necessary formulations for the micro step, restriction, macro step and matching. 

Numerical tests showed that the micro-macro method maintains good accuracy while significantly improving computational speed for both a dam break test case and a wave transport case. In our tests, the accuracy of the micro-macro method depends largely on the chosen macro model.


The work in this paper proves that the restriction and matching between models with different number of variables works well, which opens up possibilities for model variations, e.g., via different distance functions. Further research regarding the micro-macro method can consider different shallow water moment equations and switching of models. Research can be done into methods which adaptively use two different models on two parts of the domain. On parts of the domain with simpler velocity profile a less accurate model can be used compared to the parts with a complicated velocity profile. This could eventually be expanded into methods where the number of variables changes over time.

\section{References}
\printbibliography[heading = none]


\end{document}